\documentclass[a4paper,fleqn,usenatbib]{mnras}
\usepackage{newtxtext,newtxmath}
\usepackage[T1]{fontenc}
\usepackage{ae,aecompl}

\usepackage{graphicx}	% Including figure files
\usepackage{amsmath}	% Advanced maths commands
\usepackage{amssymb}	% Extra maths symbols

\title[Gravitational collapse of molecular clumps]{
Scale-free gravitational collapse as the origin
of $\rho \sim r^{-2}$  density profile -- {  a possible role of turbulence
in regulating gravitational collapse}}

\author[Guang-Xing Li]{
Guang-Xing Li$^{1}$\thanks{E-mail: gxli@usm.lmu.de (USM)}
\\
$^{1}$University Observatory Munich, Scheinerstrasse 1, D-81679 M\"unchen,
Germany\\}

\date{Accepted XXX. Received YYY; in original form ZZZ}

\pubyear{2015}

\begin{document}
\label{firstpage}
\pagerange{\pageref{firstpage}--\pageref{lastpage}}
\maketitle

\begin{abstract}
Astrophysical systems, such as clumps that form
star clusters share a density profile that is close to $\rho \sim r^{-2}$. We
 prove analytically this density profile is the result of the scale-free
 nature of the gravitational collapse.
Therefore, it should emerge in many different situations as long as gravity is
dominating the evolution for a period that is comparable or longer than the
free-fall time, and this does not necessarily imply an isothermal
model, as many have previously  believed.
To describe the collapse process, we construct
a model called the turbulence-regulated gravitational collapse model, where
turbulence is sustained by accretion and dissipates in roughly a crossing time. We demonstrate
that a $\rho \sim r^{-2}$ profile emerges due to the scale-free nature the system.
In this particular case, the rate of gravitational
collapse is regulated by the rate at which turbulence dissipates the kinetic
energy such that the infall speed {\bf can be} $20 \text{-} 50 \%$ of the
free-fall speed{(which also depends on the interpretation of the crossing time based on
simulations of driven turbulence)}.
These predictions are consistent with existing observations, which suggests that
these clumps are in the stage of turbulence-regulated gravitational collapse.
Our analysis provides a unified description of gravitational collapse
in different environments.
\end{abstract}


%%%%%%%%%%%%%%%%%%%%%%%%%%%%%%%%%%%%%%%%%%%%%%%%%%

\section{Introduction}
In many astrophysical systems, the underlying density structures are close to
spherical symmetric, which allows one to defined radial profiles where the
radially-averaged mean gas density $\rho$ is a function of the radius $r$.
In systems like star cluster-forming molecular clumps, the radial density
profile is very close to $\rho \sim r^{-2}$
\citep[e.g.][]{2002ApJS..143..469M,2003cdsf.conf..157E,2012A&A...542L..15W,2014ApJ...785...42P,2016A&A...585A.149W,2017A&A...600L..10C,2017MNRAS.469.2286Z}, and
this has became even more clear after one has constructed density profiles
of these clumps with a higher spatial dynamical range using observations obtained with different instruments/interferometer configurations \citep[e.g.][]{2017A&A...600L..10C,2017MNRAS.469.2286Z}.

The theoretical interpretation of this profile remains unclear.
% Since this
% density profile have been derived in the Singular Isothermal Sphere profile
% model
% \footnote{\url{https://en.wikipedia.org/wiki/Singular_isothermal_sphere_profile}.
% }.
 \citet[][]{1977ApJ...214..488S} studied the collapse of singular
isothermal gas spheres, and found that $\rho \sim r^{-2}$ emerges as a result of
hydrostatic balance, and during the collapse a somewhat shallower density
profile is established. Presumably,  because of historical reasons, this had
lead to the misconception that a $\rho\sim r^{-2}$ density profile must
indicate some pressure support. However, one should always bear in mind
that expected radial profile depends on many factors, including the equation of
 state as well as the initial and boundary conditions. Even after
 considering all these factors, the solutions might not be physical \citep[see e.g.][ for a thorough treatment]{1985MNRAS.214....1W}.
Another common problem with the studies during 1960-1980s was that the authors typically
assumed an isothermal equation of state, and  in systems such as
star-forming regions, gas motion is highly supersonic such that the thermal pressure is not dominant.
This makes it difficult compare observations with their model predictions.

Interestingly, the $\rho \sim r^{-2}$ profile seems to emerge
in a variety of situations where we do not expect the isothermal condition to hold.
 For example, in molecular clumps that form star clusters, the ram-pressure of
 the turbulent motion is much larger than sound velocity (with is related to
 the thermal motions of the gas). However, observations still indicate that
 a $\rho \sim r^{-2}$ profile emerges. There are also models that can
 produce the $\rho \sim r^{-2}$  profile with some free-fall
 \citep{1969MNRAS.145..271L,2015ApJ...814...48N}.
 One should explain why we still observe $\rho \sim r^{-2}$ in these very
 different conditions -- what is the cause of the $\rho \sim r^{-2}$
 profile?

In this paper, assuming a spherically-symmetric
 geometry, we present a simple scaling analysis, and propose that $\rho
 \sim r^{-2}$  profile is a natural consequence of scale-free gravitational
 collapse, where, after some time, the local infall time become linked to the
 dynamical time of the gas.
 We back up our analytical argument with an analytical model that describes
 the interplay between turbulence and gravity, and prove that a $\rho\sim
 r^{-2}$  should emerge if the kinetic energy injected into the system due to
  gravitational collapse and dissipates through turbulence.  The
 model quantitatively explains observations of these pc-scale molecular clumps.
In short, the emergence of  a $ \rho \sim r^{-2}$ profile does not necessarily
imply an isothermal model, but should be regarded as a generic feature of
scale-free gravitational collapse.

\section{Origin of the $\rho \sim r^{-2}$ profile}\label{sec:simple}
We use the method of scaling analysis \citep{1996sssi.book.....B} to derive
the expected radial profile during the gravitational collapse. We consider a
spherical gas clump. In the clump, collapse had already occurred such that within
the range between $r_{\rm in}$ and $r_{\rm out}$, the system reaches a quasi-stationary state,
where the density structure can be described as $\rho(r)$ where $r$
is the radius, $r_{\rm in}$ is the inner radius and
$r_{\rm out}$ is the outer radius. How one determines $r_{\rm in}$ and $r_{\rm
out}$ will be explained later. We are not studying how the structures are
formed, but are interested in what the structure would be had the clump
 been given enough time to relax \footnote{See \citet{2015ApJ...799..227G} for a
discussion on the initial relaxation process, where the relaxation time is
found to be comparable to the dynamical time.}.

We further impose the constraint that the mass inflow rate at $r_{\rm out}$ is
identical to the mass inflow rate at the inner boundary $r_{\rm in}$. In
reality, $r_{\rm in}$ corresponds to the radius within which gravity of the
central objects dominates, and $r_{\rm out}$ is the radius beyond
which the gas is no longer gravitationally bound
(see also \citet{2017MNRAS.465..667L} where they presented a Jeans-like formula
to derive $r_{\rm out}$). We further assume that $r_{\rm in} << r_{\rm out}$
(e.g. $r_{\rm in}$ is sufficiently small so that we can neglect the effect of  $r_{\rm in}$ on the density slope).
Due to the lack of a characteristic scale in addition to $r_{\rm out}$, we
expect the density distribution to take the form of a power-law where
\begin{equation}
\rho = \rho_0 (r/ r_0)^{- k_{\rho}} \sim r^{- k_{\rho}}\;,
\end{equation}
where  $r_0$ and $\rho_0$ are normalisation factors. The
 mass enclosed in a region of radius $r$ is
\begin{equation}
m_{\rm enclosed} \approx \int_{r_{\rm in}}^{r}  4 \pi \rho  r'^2 {\rm d } r'
\sim r^{3 - k_{\rho}}\;.
\end{equation}

At radius $r$, the radial acceleration is
\begin{equation}\label{eq:accel}
a_{\rm r} \approx  G m_{\rm enclosed} / r^2\sim r^{1 - k_{\rm
\rho}}\;,
\end{equation}
 the virial velocity  (velocity at which the systems is virialised) is
\begin{equation}
v_{\rm ff} \approx \sqrt{  G m_{\rm enclosed} / r} \sim r^{1- k_{\rho
}/2}\;,
\end{equation}

and the free-fall time is
\begin{equation}
t_{\rm ff}\approx \sqrt{1/ G \rho} \approx r / v_{\rm ff}  \sim r^{k_{\rho
}/2}\;.
\end{equation}

When gravity is dictating the evolution of the system, the infall velocity at
radius $r$ must be proportional to the free-fall velocity $v_{\rm infall}
 \sim   f v_{\rm ff}$ \footnote{  Here, it is not straight forward to derive
 the value of $f$ directly. It is easier to consider some simple cases: For free-fall
 collapse $f$ can approach and even exceed unity and otherwise $f  \lesssim 1$.
 In Sec.
 \ref{sec:model}, we will demonstrate how turbulence can regulate the collapse
 to $f < 1$, where value of $f$ determined
 by the energy dissipation efficiency of the turbulence.}. Our clump has a mass
 inflow rate of
\begin{equation}\label{eq:infall}
\dot m \approx 4 \pi r^2 \rho v_{\rm infall}\approx f r^2 \rho v_{\rm ff} \sim
r^{3- 3 k_{\rho }/2}\;.
\end{equation}

To drive the radial density profile of such a self-similar infalling system, we
{\it demand} that the mass inflow rate $\dot m $ to be a constant ($\dot m \sim
r^0 =\rm  Constant$ ) at different radii. Using Eq. \ref{eq:infall}, we have
\begin{equation}
3 - 3 k_{ \rho} / 2 =0  \rightarrow k_\rho = 2, {\rm thus,}\; \rho\sim r^{-2}\;,
\end{equation}
which explains the $r^{-2}$ profile.
 Finally, we remind the reader
that at a given radius $r$, $v_{\rm infall} \sim v_{\rm ff}$ automatically
implies that the infall time $t_{\rm infall} = r / v_{\rm infall}$ is positively
related to the free-fall time $t_{\rm ff} = r / v_{\rm ff}$.

We briefly explain what determines $r_{\rm in}$:
From Eq.
\ref{eq:accel}, assuming $\rho \sim r^{-2}$, we expect a radial
gravitational acceleration that scales with $r^{-1}$.
A necessary condition for our analysis to be applicable is that the
 acceleration estimated by Eq. \ref{eq:accel} should be accurate there.
In reality (e.g. the case of star cluster-forming clumps), at very small radii,
gravity from the central proto star-cluster dominates, and we expect our
results to be accurate beyond a given radius where the deviation of
acceleration from what one expects from Eq.
\ref{eq:accel} is not significant (as has been discussed in
\citet{2015ApJ...804...44M}).

To sum up, we have demonstrated that a spherical system that undergoes
self-similar infall tend to relax to $\rho \sim r^{-2}$  during a self-similar
infall, if (1) the system is gravitationally bound (2) the system has sufficiently relaxed and has
reached a quasi-stationary state, (3) the amount of mass is roughly conserved
during the infall (e.g.
the gas consumption due to star formation during the infall is negligible), and
(4) the mass of the central object does not cause the acceleration to deviate from the one
predicted by Eq. \ref{eq:accel} by much.

These conditions are satisfied at least approximately in many
situations, and can explain the $r^{-2}$ profile found in previous papers
\citep[e.g.][]{1969MNRAS.145..271L,1969MNRAS.144..425P,2015ApJ...814...48N}\footnote{Where
the analytical result of \citet{1969MNRAS.144..425P} is $\rho \sim r^{-12/7}$,
and $12/7$ is almost $-2$. }.
What is universal among these models is that after some
relaxation processes, the infall time of the gas become linked to the free-fall
time of the gas, and this leads to the $\rho \sim r^{-2}$ profile.
However, it remains to be determined  what influences the infall
speed.
In the next section, we construct a simple analytical model that involves the
interplay between turbulence dissipation and gravitation collapse, and
demonstrate that the ratio between the infall speed and free-fall speed is
related to the ratio between the turbulence dissipation rate and gravitationally
collapse rate.

\section{Turbulence regulated gravitational collapse of molecular clumps}
\label{sec:model}
We establish a concrete physical model to  describe how the
interplay between turbulence and gravity gives rise to a $\rho \sim r^{-2}$
profile, and determine the rate of gravitational collapse. The model shares a
common view with many of the previous ones, e.g.
\citet[][]{2015ApJ...804...44M,2016A&A...591A..31L,2017MNRAS.465..667L}.

We consider a clump with a density profile $\rho(r)$ and
an infall velocity profile $v_{\rm infall}(r)$.
In essence, the structure of
molecular interstellar medium is complicated, and the close-to-spherical
geometry which we adapt is an approximation, which we believe is a reasonable
one: Observationally, these structures are indeed close to be spherical. One
good example is the well-resolved nearby star-cluster forming region NGC1333.
These observations thus provide an empirical justification to our assumption.
Theoretically, combined with some flow asymmetry, gravity is believed to be able
to drive collapse away from sphericity, producing sheets, and then filaments
\citep{1965ApJ...142.1431L,1970A&A.....5...84Z}. This mechanism should explain the existence of filamentary structures in ISM \citep{2014ApJ...791..124G}.
 However, it has been later shown that the effect can be weakened by
the presence of a  steep density gradient
\citep{1983MNRAS.203..265G,2000ApJ...535..402L}.
Since our
clumps are centrally-condensed, we expect the density gradient
stabilisation to be effective, thus the deviation from sphericity due to gravity
should not be severe for our case. Besides, a turbulent pressure can also
help to reduce the radial density asymmetry. These justify our spherical
assumption.

We also assume that the profiles are  are independent of time
$t$.
This corresponds to the situation where there is a continuous mass inflow starting from $r_{\rm
out}$ to $r_{\rm in}$. When mass gets
accumulated at the centre, we expect $\rho$ to increase also with the time. In
these cases, our analysis is still valid as long as the time at which $\rho$
increases is much longer than the dynamical time of the system.

The evolution of the clumps is dominated by a combination of
turbulence and gravity. Gravity is responsible for driving the turbulent
motion \citep{2008MNRAS.385..181F,2010A&A...520A..17K,2012ApJ...750L..31R,2015ApJ...804...44M},
and the turbulence motion decays within a few crossing times
\citep{1998ApJ...508L..99S,1998PhRvL..80.2754M}.
The role of turbulence is to dissipate the energy generated by gravitational at
different scales.

We assume that at radius $r$, the turbulent velocity dispersion
$\sigma_{\rm v}$ is determined by the virial velocity $\sigma_{\rm v, vir}$. This corresponds to
the case where the clumps are supported by the virialised quasi-random motions
which are generally assumed to be turbulence, and accretion is the major source
of the turbulence
\citep{2008MNRAS.385..181F,2010ApJ...712..294E,2010A&A...520A..17K}.
Previously, \citet{2015ApJ...804...44M} carried out a detailed analysis of gravitational collapse in a turbulent medium using a 1D model.
In their model, turbulence is driven by adiabatic heating
\citep{2012ApJ...750L..31R}. Here, we assume that at different radii, gas in the clump is virialized.
Our view is
supported by the fact that observationally, these clumps are found to
be virialized over multiple scales\footnote{See e.g.
\citet{2015A&A...578A..97L} where they pointed out that the value of the virial
parameter is related to how one selects regions. If the regions are properly
defined, all the regions in the Persues and Ophiuchus that they studied are
gravitationally bound. This is consistent with previous results
, e.g. \citet{2009Natur.457...63G}.}, as a theoretical treatment, it is
appropriate to invoke a virialised turbulence model and remove the dependence of
the results on the initial turbulence seed assumed in the previous model.
{Due to the presence of the infall,  gas in the
clumps can never establish a complete virial equilibrium: if the infall time is
comparable to the dynamical time, infall motion can not be neglected; if
the infall time is longer than the dynamical time, the infall motion is slow and
can be neglected as a first-order approximation, where the gas can stay close
to the virial equilibrium. In observations, the infall speed is a
fraction of the free-fall speed
\citep[e.g.][]{2003cdsf.conf..157E,2012A&A...542L..15W,2016A&A...585A.149W},
thus the our assumption of virial equilibrium is still valid given the presence
of the infall. }

%
% {Therefore, the condition which we assume in our model is at least achievable
% through adiabatic heating, although we believe that other mechanism might also contribute to
% the virialisation of the turbulent flow.}

% Our model neglects the star formation

% process, and captures the interplay between turbulence and gravity. The model is
% valid outside $r_{\rm in}$ which in which the star formation process and the
% contribution of gravitational potential from stars becomes important.
In a real molecular clump, star formation has some effects on
the gas evolution. Generally speaking, star formation has two effects: First,
it consumes mass, and second, it provides feedback. Because it
is believed that the star formation efficiency is a strong function of gas
density \citep{2014prpl.conf...77P}, it is only able to consume gas efficiently
at the very inner part of the clumps and
acts as a mass sink, and this has already been properly considered in our model.
Star formation also produces feedback. In general, the feedback effects include
wind-driven molecular outflows, stellar winds and supernova explosions.
The feedback from wind-driven molecular outflows is mostly contributed from
low-mass stars.
These outflows might drive turbulence
\citep{2006ApJ...640L.187L,2007ApJ...659.1394M}. However, the effects are
considered to be relatively weak \citep{2015ApJ...815...68M}.
In essence, the efficiency of the outflow feedback depends on the outflow
geometry.
According to some more recent models, the outflowing gas is contained in a thin
layer \citep{2013A&A...559A..23L}, the coupling between the
outflow and the environment is inefficient and the impact of these outflows on
the large-scale clump dynamics can be neglected.
Wind from massive stars and supernova explosions can be very powerful and will
disrupt the clumps at some stage. Here, we are only interested in the clump
structures before the disruption occurs. Thus, in spite of its simplicity, our
model should be able to capture the general behaviour of the gas flow in the
observed molecular clumps.

\subsection{Parameterization}

We parameterize the density
structure as
% \footnote{Here, $\rho_0$ have a unit of
% $M L^{3 - k_{\rho}}$ where $M$ is mass and $L$ is size.}
\begin{equation}\label{eq:rho}
\rho(r) = \rho_0 \Big{(}\frac{r}{r_0}\Big{)}^{-k_{\rho}}\;,
\end{equation}
The infall velocities at different radii are parameterised as
% \footnote{Here,
% $v_0$ have a unit of $L^{1 - k_{\rm v}} T^{-1}$ where $L$ is
% size and $T$ is time.}
\begin{equation}
v_{\rm infall}(r) = v_0 \Big{(}\frac{r}{r_0}\Big{)}^{-k_v}\;.
\end{equation}
where $v_0$ is a normalisation parameter whose value will be determined later.

We work with Eulerian coordinates. At $(r, r + {\Delta}r)$, the shell have a
mass of
\begin{equation}
m_{\rm shell} = 4 \pi r^2 \rho(r) {\rm d} r = 4 \pi\rho_0 r_0^2
\Big{(}\frac{r}{r_0}\Big{)}^{2-k_{\rho}}{\Delta}r\;,
\end{equation}
and the mass enclosed in such a shell is
\begin{equation}
m_{\rm enclosed} = \frac{4 \pi\rho_0 r_0^3}{3 -
k_{\rho}}\Big{(}\frac{r}{r_0}\Big{)}^{3-k_{\rho}}\;.
\end{equation}

We assume that  gas in each shell is approximately virialized,
\begin{equation}
\sigma_{\rm v, 3D} \approx \sqrt{\frac{G m_{\rm
enclosed}}{r}}\;,
\end{equation}
where $\sigma_{\rm v,3D}$ is the three-dimensional velocity dispersion \footnote{
Here, we have chosen to use the 3-dimensional velocity dispersion, because we will be
estimating the turbulent energy dissipation rate. For convenience, the formalism
has been chosen to be  consistent with simulations such as
\citet{1998ApJ...508L..99S}, where the energy dissipation rate is estimated with
respect to $l/\sigma_{\rm v}^{\rm 3D}$, $l$ is the {\bf cloud} size and
$\sigma_{\rm v}^{\rm 3D}$ is the three-dimensional velocity dispersion.  }.
The crossing time can be estimated as \footnote{Here, we must remind
the reader that the crossing time might be defined differently in different models, and in our
case the crossing time is defined as the ratio between the turbulence driving
scale and the 3-dimensional velocity dispersion at that scale, which is
consistent with previous works e.g.
\citet{1998ApJ...508L..99S,1998PhRvL..80.2754M,1999ApJ...524..169M}.}
\begin{equation}\label{eq:tcross}
t_{\rm cross} \approx \frac{2\times \beta \times r}{\sigma_{\rm v,3D}}  \approx  2 \times \beta\
\times \Big{(}\frac{r^3}{G
m_{\rm enclosed}(r)}\Big{)}^{1/2}\;,
\end{equation}
 where the factor of 2 comes from the fact that the diameter is two times
the radius. $\beta$ is introduced as the ratio between the driving scale of the turbulence and the clump diameter. If the driving scale of the turbulence is comparable to the clump diameter $\beta = 1$; if the driving scale of the turbulence is much smaller, $\beta < 1$ \citep[details concerning the driving of turbulence has been described in][]{1998ApJ...508L..99S}. In this paper, we assume that the driving scale of the turbulence is comparable to the clump diameter, where $\beta = 1$. Note that the value of $\beta$ is also dependent on the interpretation of the crossing time in the simulations. Our referee pointed out that Fourier driving involves sine waves that contains both positive and negative velocities, and thus the actual crossing time for a mode of wavelength equal to the box size may be the time necessary for crossing only half of the box length. We have chosen $\beta =1$, where the driving scale is comparable to the cloud size. Our assumption is consistent with the discussion in \citet{{1998ApJ...508L..99S}}.
The kinetic energy of such a shell is
\begin{equation}
E_{\rm kin} \approx  \frac{1}{2} m_{\rm shell} \sigma_{\rm v}^2 = \frac{1}{2}
\frac{G (4 \pi \rho_0 r_0^2)^2}{3 -k_{\rho}} \Big{(}\frac{r}{r_0}\Big{)}^{4 - 2
k_{\rho}} {\Delta} r\;.
\end{equation}
 { We note that although the shell thickness $\Delta r$ is introduced here
 as a parameter, we are interested in the radially-averaged means, and our
 holds exactly as $\Delta
 \rightarrow 0$. }

\subsection{Mass and energy conservation in a clump}
We derive the clump structure using mass and energy
conservation.
Because of mass conservation, the mass inflow rate $\dot m = 4 \pi r^2 \rho
v_{\rm infall}$ should stay constant at different radii, thus
\begin{equation}
\dot m = 4 \pi r^2 \rho v_{\rm infall} = 4 \pi r^2 \rho_0
\Big{(}\frac{r}{r_0}\Big{)}^{-k_\rho} \Big{(}\frac{r}{r_0}\Big{)}^{-k_v} \sim
r^{2 - k_\rho - k_v } =\rm Constant\;,
\end{equation}
which gives
\begin{equation}
2 - k_\rho - k_v = 0\;.
\end{equation}

The rate of gravitational energy release from infall is
\begin{eqnarray}\label{eq:e:gravity}
\dot E_{\rm gravity} &=&   \frac{G m_{\rm
enclosed}m_{\rm shell} v_{\rm infall}}{r^2} \nonumber \\
&=& \frac{(4 \pi  \rho_0)^2 G v_0 r_0^3}{3 - k_{\rho}}
\Big{(}\frac{r}{r_0}\Big{)}^{3- 2 k_{\rho}-k_v}{\Delta}r\;,
\end{eqnarray}

The turbulent energy of such a shell is expected to decay within $n$ (which is
believed to be around 0.8,
\citet{1998ApJ...508L..99S,1998PhRvL..80.2754M,2009ApJ...691.1092L})  crossing
times, which
gives an energy dissipation rate
\begin{equation}\label{eq:dissipation}
\dot E_{\rm kin} \approx \frac{E_{\rm kin}}{n \times t_{\rm cross}} =
\frac{ G^{\frac{3}{2}} (4 \pi \rho_0)^{\frac{5}{2}}r_0^4 }{ 4 n \beta (3 -
k_{\rho})^{\frac{3}{2}}} \Big{(}\frac{r}{r_0}\Big{)}^{4 -
\frac{5}{2}k_{\rho}}{\Delta}r\;.
\end{equation}

We require that energy conservation holds exactly for each shell, and this gives
\begin{equation}
\dot E_{\rm kin}(r) =  \dot E_{\rm gravity}(r)\;,
\end{equation}
which implies (using Eq. \ref{eq:e:gravity} and Eq. \ref{eq:dissipation})
\begin{eqnarray}\label{eq:ks}
k_\rho &=& 2 \\ \nonumber
k_{\rm v} &=& 0    \\ \nonumber
 v_0 &=&(4 \pi G \rho_0 r_0^2)^{\frac{1}{2}}\times
 \frac{1}{ 4 n \beta }\;.
\end{eqnarray}

Observationally, $r_{\rm clump}$ and $M_{\rm clump}$ can be measured. We can
use these values to determine the normalisations of Eq. \ref{eq:rho},
\begin{equation}
\rho_0 = \frac{1}{4 \pi r_0^2}\frac{M_{\rm clump}}{r_{\rm clump}}\;,
\end{equation}
and \begin{equation}\label{eq:v:infall}
v_{\rm infall} \approx \Big{(} \frac{G M_{\rm clump}}{r_{\rm
clump}}\Big{)}^{1/2}\times
\frac{1}{4 n \beta} \;.
\end{equation}
If one defines the
 free-fall velocity of such a clump as $v_{\rm ff}
\approx \sqrt{G m_{\rm clump}/ r_{\rm clump}}$, the predicted infall velocity is
always a fraction ($1/ 4 n \beta$) of the free-fall
velocity, where $\beta$ is determined by turbulence driving. We assume that the driving scale is comparable to the clump diameter, such that $\beta = 1$ (see discussions followed Eq. \ref{eq:tcross}).
%
% \footnote{One can also understanding this using the formalism
% presented in \citet{2012ApJ...750L..31R} where turbulence is driven
% through adiabatic heating. To estimate the limiting velocity,
% one can use their equation 7 and let ${\rm d} v /{\rm d a} = 0$, which gives
% $v_{\rm infall} \sim 1/ 2n v_{\rm turb}$. To verify this, note that by
% definition, their contraction speed {  (which is defined on the box as a
% whole)} is twice our infall speed ({  which is defined with respect to the
% box center}).} .
 Although the general conclusion that that an inefficient energy
dissipation will lead to an a slow gravitational collapse is true, the above
derivation has a couple of uncertainties. One noticeable one is the
turbulence energy dissipation rate estimate. Here we are applying the energy
dissipation rate estimated in simulations such as
\citet{1998ApJ...508L..99S,1998PhRvL..80.2754M,1999ApJ...524..169M} to our model. Using the
simulation results leads to two issues: first, the physical condition in the
numerical simulation has to be identical to the physical condition of the clumps, and second, the crossing
time defined in the simulations and in observations must be comparable.  The
simulations have parameters that are chosen to resemble that of the molecular
ISM, and we expect the results from the simulation to be accurate in this
sense. However, in the clumps, the presence of a radial density
gradient might physically affect the energy dissipation process and change the
dissipation rate. How strong this effect is remains undetermined. Another
uncertainty comes
from the way one defines the crossing time:
The crossing time formula involves a scale and the velocity
dispersion at that scale. In both the simulations and our model, the velocity
dispersion is the three-dimensional velocity dispersion.
Therefore the velocity dispersion estimate is not a major source of uncertainly
\footnote{In the text of
\citet{1998ApJ...508L..99S,1998PhRvL..80.2754M,1999ApJ...524..169M}, the author did not explicitly specify whether they have used 1D or 3D velocity
dispersion. However, the kinetic energy densities are estimated using the
formula $1/2 \rho \sigma_{\rm v}^2$ instead of $3/2 \rho  \sigma_{\rm v}^2$,
suggesting that the authors are using the 3D velocity dispersion. }.
However, it is not straightforward to find a unique way to compare $l$:
in our formalism we have assumed that at radius $r$, the relevant scale
is simply the diameter $2 r$, and in the simulation of
\citet{1998ApJ...508L..99S,1998PhRvL..80.2754M}, $l$ is {the driving scale.
In other words, one of the  assumption in our estimate of turbulence
energy dissipation rate is we have assumed that the driving
scale of the turbulence is comparable with the clump radius. Although this is
plausible, it is is still dependent on the nature of turbulence in the
clumps, which is not fully clear. }

\subsection{Universality of the $\rho\sim r^{-2}$ profile }
The turbulence-regulated gravitational collapse model provides us an
excellent example to understand the relevance of the analytical arguments presented in Sec.
\ref{sec:simple} to real systems.
First, in  our analytical study, we only made use of the scale-free
nature of gravitational collapse, and here we take the interplay between
turbulence and gravity into account. Since  the explicit inclusion of energy
dissipation rate estimates due to equilibrium cascade
in the model does not introduce additional scales, we recover our $\rho\sim
r^{-2}$ profile expected in the previous section. Second, by writing down
detailed equations describing the interplay between gravitational contraction and turbulence dissipation in
a centrally-condensed clump, we are able to analytical derive the relation
between turbulence dissipation efficiency and infall: the infall velocity
derived in Eq. \ref{eq:v:infall} is slower than the free-fall velocity by a
factor of $4\times n$. In our clumps, the infall velocity is slower than the
free-fall velocity because the turbulence dissipation time $t_{\rm dissipation}$ is longer
than what it required to be if the system undergoes free-fall. To put it in a
different from:
\begin{equation}
\frac{v_{\rm infall}}{v_{\rm ff}} \approx  \frac{1}{4}
\frac{t_{\rm dissipation}}{t_{\rm crossing}}\;,
\end{equation}
where the rate of gravitational collapse is linked to the efficiency of
turbulence dissipation. The collapse is slow in turbulence-supported clumps due
to the inefficiency of energy dissipation.
%
% \footnote{ In \citet{2015ApJ...804...44M}, the author attributed the slowness of the collapse
% found in their models to some pressure support. We believe that the energy
% dissipation rate plays a more fundamental role than pressure.
% According to our formalism, it is the inefficiency of turbulence dissipation
% that leads to the slowness of gravitational collapse.} .

\section{Observational correspondence}
\subsection{$\rho \sim r^{-2}$ profile in star-cluster forming clumps}

The $r^{-2}$ profile has been widely seen in observations of star
cluster-hosting clumps
\citep[e.g.][]{2002ApJS..143..469M,2003cdsf.conf..157E,2012A&A...542L..15W,2014ApJ...785...42P,2016A&A...585A.149W,2017A&A...600L..10C,2017MNRAS.469.2286Z}.
Observations also found that the infall speed is a fraction of the free-fall
speed \citep{2003cdsf.conf..157E,2016A&A...585A.149W,2012A&A...542L..15W}. How
well can these be explained by existing theories?

Quite a few models can explain a $\rho \sim r^{-2}$-like density profile. There
are two categories of models. The first category includes those
without infall. The most famous one is the \citet{1977ApJ...214..488S} solution
where the author attributed the profile to the fact that the system must achieve
hydrostatic equilibrium before the collapse starts. The other set of models
involves infall.
These include models that involve free-fall collapse \citep{1969MNRAS.145..271L,1969MNRAS.144..425P,2009ApJ...707.1023V,2011MNRAS.411...65B,2014ApJ...781...91G,2015ApJ...814...48N,2017arXiv170802401D},
where the $r^{-2}$ profile is found to be the analytical solution (such as the
case of
\citep{1969MNRAS.145..271L,2015ApJ...814...48N,2017arXiv170802401D}), as well as models that involve gravity and turbulence,
including the turbulence-regulated gravitational collapse model presented above.
% A similar study has been presented in  \citet[][]{2015ApJ...804...44M} where the
% turbulence is adiabatically driven, and the density profile at large
% radii is $r^{-1.8}$, very similar to our results.

 Our general analysis has suggested that the $\rho \sim r^{-2}$ profile is a
 generic signature of scale-free gravitational collapse, and this is independent
 on whether the collapse is free-fall or not. After accepting the analysis, it
 is no longer a surprise that a universal profile has emerged in these different settings. This has two implications for the
 interpretation of observations: first, the $\rho \sim r^{-2}$ profile do not necessarily imply
 some pressure support, as \citet{1977ApJ...214..488S} might had believed. It
 emerges in models that involve free-fall collapse and turbulence-regulated gravitational
 collapse as well.
 Second, even if a clump is found to be collapsing, based on the $r^{-2}$
 profile, it is difficult distinguish between a pure free-fall and a
 turbulence-regulated collapse.

% These different infall models can be distinguished using the infall velocity.
% Previously,  it has been argued that the clumps might exist for a few
% crossing timescales, and their physical state might approach some sort of equilibrium
% \citep{2003ApJ...585..850M,2006ApJ...641L.121T}. On the hand, it has been
% proposed that turbulence plays little role in these clumps, and the clumps are
% simply objects that undergo multi-scale free-fall
% \citep{2009ApJ...707.1023V,2011MNRAS.411...65B,2015ApJ...814...48N}.
%  Our analysis suggests that in both cases, the characteristic $r^{-2}$ profile can be produced if gravity dominates, and
%  this is in agreement with the analysis presented in Sec.
% \ref{sec:simple}. One should also note that from this density profile alone, it
% is difficult to tell if the collapse is turbulence-regulated or free-fall. To resolve this ambiguity one should look at velocity structure of these clumps.

To further distinguish these infall models, we should look at the infall
velocity. According to Sec. \ref{sec:model}, if turbulence is regulating the
collapse, we expect to observe an infall velocity that is much smaller than the
free-fall velocity. This is very different from the
free-fall models where the infall velocity should be close to the
free-fall velocity. In observations,  the velocities are found to be around
$\sim 5 - 30 \%$ of the free-fall velocity
\citep[e.g.][]{2003cdsf.conf..157E,2012A&A...542L..15W,2016A&A...585A.149W},
suggesting that some mechanism is regulating the collapse.

According to Sec. \ref{sec:model}, if turbulence dissipates in $n$ crossing
times. Assuming that the driving scale of turbulence is comparable to the clump diameter, $\beta\approx 1$ (Eq. \ref{eq:tcross}),
the infall velocity should be roughly $1/4 n$ times the free-fall
velocity. The turbulence in the
clumps is believed {  to dissipate
within around $n= 0.7$ crossing time (for a Mach 7 hydrodynamic turbulence)
 to $n=1.2$ crossing time (for a Mach 1.2
hydrodynamic turbulence, according to \citet{2009ApJ...691.1092L}). This would
imply that the infall speed should be roughly $1/4 n \approx 20 \%$ to $31\%$ of the free-fall speed. This
is comparable to the infall speed seen in observations (which is $\sim 5 - 30
\%$ of the free-fall velocity) \footnote{  Note that although including a magnetic field can also change the energy
dissipation rate thus the infall speed, the effect should not be not particular
strong.
E.g.
according to \citet{2009ApJ...691.1092L}, a strongly magnetized turbulence with $\beta =
0.02$ has $n = 1.7$. But such kind of flow has not been seen observationally.
Typically, the energy density of the gravitational field is comparable to the
energy density of the magnetic field
\citep{2010ApJ...725..466C,2015Natur.520..518L,2017arXiv171102417L}, making
such a small $\beta$ unlikely.}. Thus the observations and our
turbulence-regulated gravitational collapse model is largely consistent.
However, one needs to bear in mind that there are sources whose infall velocity
is too small to be explained by turbulence regulation.  These sources might
have a different nature, and this issue remains to be understood.}
In general, gravitational collapse regulated by turbulence dissipation can
quantitatively explain the slowness of gravitational infall.

\subsection{Connection between $\rho\sim r^{-2}$ profile and the surface density
PDF}
A convenient way to quantify a structure is to derive the density PDF
(Probability distribution function).
Since the density profile in our model is a power-law, it is natural to expect
the clumps to have power-law density PDFs (as
has been discussed in \citet[][]{2011ApJ...727L..20K}) \footnote{Where the
authors proposed that $P(\Sigma) \sim \Sigma^{-2}$  represents an attracting solution to the
system.}.
It is indeed the case: the $\rho \sim r^{-2}$ profile implies a surface density PDF of
$P(\Sigma) \sim \Sigma^{-2}$ where $\Sigma$ is surface density
\citep{2014ApJ...781...91G,2016MNRAS.461.3027L} \footnote{In general, a radial
profile of $\rho \sim r^{- \gamma_{\rho}}$ implies a surface density PDF of
$P(\Sigma) \sim \Sigma^{2 / (1 - \gamma_\rho)}$.}, and interestingly, this is
consistent with fact that in observations \citep[e.g.][]{2009A&A...508L..35K,2015A&A...576L...1L,2016MNRAS.461.3027L}.

\section{Conclusions}
The $\rho\sim r^{-2}$ profile has been seen in many self-gravitating systems,
and the interpretation of this profile has been ambiguous. In many cases, these
profiles are interpreted using an isothermal solution or free-fall collapse. In
this paper, based on a simple scaling analysis, we argue that this profile can be created in a wide
range of situations where, after some initial relaxation processes, the time for gravitational
collapse is linked to the local free-fall time. When this
condition has been reached, the collapse will occur in a self-similar fashion,
which leads to the $\rho\sim r^{-2}$ profile. This explains why the $r^{-2}$
profile has been found in simulations of different settings.

%  The
% analytical arguments are general, and the results are
% independent on whether the collapse is free-fall or it is regulated by other
% processes such as turbulence, as long as the regulating processes do
% not introduce additional scales.

We further develop an analytical model called turbulence-regulated
gravitational collapse model that incorporates the interplay between turbulence
dissipation and driving of turbulence by gravitational collapse.
Using constrains such as mass and energy conservation, we find that the
stationary solution should exhibit the $\rho\sim r^{-2}$ profile. The infall
speed should be a fraction of the free-fall speed, and this is mainly caused by
the fact that the rate of gravitational collapse is regulated by the rate at
which turbulence dissipates the kinetic energy
($v_{\rm infall}/v_{\rm ff} \approx  1/4 \times
t_{\rm dissipation} / t_{\rm crossing}$, {  where, according to
\citet{2009ApJ...691.1092L}, $t_{\rm dissipation} \approx 0.7 \text{-} 1.1 \,
t_{\rm crossing}$})
Turbulence dissipation is able to regulate the gravitational
collapse  such that the infall speed is $20 \text{-} 30 \%$ the free-fall
speed, although this is still dependent on the driving of turbulence, as can be seen in Eq. \ref{eq:tcross} and Eqs. \ref{eq:v:infall}. This turbulence-regulated gravitational collapse model roughly explains the
range of infall speed observed for parsec-scale molecular clumps in the Milky
Way.

 In essence, the density profile is a reflection of what physical
processes are influencing the evolution, where the emergence of the $\rho\sim
r^{-2}$ profile is an unique signature indicating that gravity is dominating the evolution of a system.
The observed infall speed contains information concerning how the collapse is
regulated:
when the system
undergoes free-fall, the infall speed is merely the free-fall speed, and when turbulence is regulating the collapse, the infall is
slower because the rate at which turbulence can remove kinetic energy is
much lower than the rate of release of gravitational energy during a free-fall
collapse.
% The collapse is somewhat slow as turbulence can not dissipation
% energy that fast.
% {   The turbulence-regulated
% gravitational collapsed model can quantitative explain the observed infall speed. However, the
% predicted infall speed in still larger than the observed values, and the
% discrepancy can either be caused by the use of the energy dissipation law in our
% model, as well as a lack of understanding of turbulence dissipation in the
% clumps.}
% , and in essence, the slowness of turbulence dissipation is thus
% related to the slowness of gravitational collapse.
% We note that this behaviour has also been
% found in self-gravitating granular systems, where the the same $\rho\sim r^{-2}$
% emerge because of the scale-free nature of the system.
The analysis provides
a unified vision into gravitational collapse in different systems.\\

\noindent {\it Some final notes from the author:} 1. Dr. Y-Q Lou pointed out that apart from the anabatic heating, there are numerous instabilities through which turbulence can be generated \citep{2012MNRAS.420.2147L}. 2. Similar to our Sec. \ref{sec:simple}, \citet{2017MmSAI..88..533B} is another example where insights can be obtained by exploiting the continuity equation. 3. A discussion with Prof. S. Donkov suggests that similar to our case, the $\rho\sim r^{-2}$ profile can be achieved with a combination of \citep{2017arXiv170802401D}.

 % \citet{2017MmSAI..88..533B} proposed another interesting approach to drive the structure of the ISM using mass conservation.

 \section*{Acknowledgements}

 Guang-Xing Li is supported by the Deutsche
Forschungsgemeinschaft (DFG) priority program 1573 ISM- SPP. We thank the
referee whose careful reviews and insightful comments which greatly improved the paper.

%%%%%%%%%%%%%%%%%%%%%%%%%%%%%%%%%%%%%%%%%%%%%%%%%%



%\bibliographystyle{mnras}


%%%%%%%%%%%%%%%%%%%%%%%%%%%%%%%%%%%%%%%%%%%%%%%%%%

%%%%%%%%%%%%%%%%% APPENDICES %%%%%%%%%%%%%%%%%%%%%

\bibliography{paper}

%%%%%%%%%%%%%%%%%%%%%%%%%%%%%%%%%%%%%%%%%%%%%%%%%%

% Don't change these lines
\bsp	% typesetting comment
\label{lastpage}
\end{document}